\documentclass[twocolappendix,  twocolumn]{openjournal}

 \usepackage{amssymb}
 \usepackage{natbib}
\usepackage{graphicx}

\usepackage{xcolor}
 
\usepackage[utf8]{inputenc}
\usepackage[english]{babel}
\usepackage{hyperref}
\hypersetup{
    colorlinks=true,
    linkcolor=blue,
    filecolor=blue,      
    urlcolor=red,
    citecolor=blue,
}
\usepackage{color,colortbl}

 \DeclareGraphicsExtensions{.bmp,.png,.jpg,.pdf}
\usepackage{verbatim}
\usepackage[normalem]{ulem}
\usepackage{orcidlink}
\usepackage{soul}

\begin{document}

    \title{Evidence for large scale  compressible turbulence in the ism  of  CSWA13, a 
star-Forming Lensed Galaxy at z = 1.87 with outflowing wind}
\shorttitle{outflow  compressible turbulence}
\author{Itzhak Goldman $^{1,2}$}
\affiliation{$^{1}$Physics Department, Afeka College, Tel Aviv 6998812, Israel}
\affiliation{$^{2}$ Astrophysics Department, Tel Aviv University, Tel Aviv 6997801 , Israel}

\begin{abstract}
  Recently,  \citet{Vasan+2024}presented spatially  resolved observations of  a wind outflow in CSWA13, a  gravitationally lensed Star-Forming galaxy at $z = 1.87$. The gravitational lensing allowed for a substantially improved spatial and kinematic resolution of the wind and of the nebular gas.  In this paper we take advantage of the  resolved data  to test for the existence   of turbulence and to study its nature.  
  
  We obtained the autocorrelation functions of the two velocity fields, and   derived the  spatial structure functions  of the residual nebular and wind velocities along the major axis of the galaxy. The structure functions, of both velocity fields,  reveal the existence of an underlying $k^{-2}$  power spectrum scaling. This scaling suggests the existence  of  supersonic compressible  turbulence. The autocorrelation functions exhibit correlations over scales comparable to the total extent of the velocity fields. Thus, the turbulence is a large scale one. The turbulent timescale corresponding to the largest scale is about 200 Myr,  an order of magnitude larger than the estimated age of the   wind and of the young stars. This implies   that the turbulence  in the ism formed well before the wind and the young stars. Given the large spatial scale of the turbulence,  it is plausible that the source of the turbulence is a large scale one e.g.  a merger or  tidal event that triggered the formation of molecular clouds, in the cores of which,   the young stars formed. A steepening of the  structure functions on the  smaller  scales provides an estimate   for the effective depth  along the line of sight  of the turbulent layer. The latter    turns out to be  $\sim 2 kpc$. 
\end{abstract}
 \maketitle

\section{introduction}

High redshift galaxies are characterized by high star formation rates as well as  outflowing winds, generated by the young stars or by AGNs, e.g. \citep{Bournaud+2009, Hoffmann+2022, Rizzo+2021, Sanders+2023, Shah+2022}.  The high rate of star formation is attributed to the assembly process of the galaxy, e.g. \citep{Sanchez+2014, Bennett+Sijacki2020, Putman2017}. The gas supply can be in the form of  inflow from the circum galactic medium (CGM) and also by more violent events  such as   mergers and tidal interactions.   

 Observations of high redshift galaxies  display velocity dispersions that are usually interpreted as manifestation of turbulence e.g. \citep{Stark+2008, Burkert+2010}. It has been argued that accretion onto disk galaxies can generate large scale turbulence, in particular at the disk outskirts, e.g. \citep{ Forbes+2023, Goldman+Fleck2023}. Turbulence can be generated also by mergers and tidal interactions. To establish the existence of turbulence and moreover, to understand its nature, a power spectrum or structure function  of the velocity field are  needed. This in turn, demands observations with high enough spatial resolution which, for galaxies at high redshifts, are challenging.
 
Gravitational lensing can help with this regard.     A  recent paper   \citep{Vasan+2024} presented a study of a wind outflow in CSWA13,  which is   a gravitationally lensed star-forming galaxy at $z = 1.87$. The gravitational lensing allowed for a substantially  improved spatial and kinematic resolution. The   authors
     obtained, among other results,  two velocity fields along the major axis of the galaxy:  the nebular gas velocity traced by the $C_{|||]}$ emission line, that represents also the velocity of the young stars embedded in the nebular gas, and the wind velocity traced by the $Si_{||}^*$ florescent emission  line. Each of these velocity fields , exhibits a large scale shear.  
     
  In the present paper we set to check wether these velocity fields can be used to test for the existence of turbulence, and if so to obtain its characteristics.
  
The two residual velocity fields are obtained in section 2. The autocorrelation functions are presented in section 3. In section 4  we obtain the structure 
functions. Discussion is presented in section 5.
 In   Appendix A, the theoretical structure function of a quantity that is the  result of integration along the line of sight direction, is derived. This is used to provide an estimate of the  depth of the turbulent layer. In Appendix B, the power spectrum of a quantity that is the  result of integration along the line of sight direction, is derived.

\section{The residual velocity fields}  
  
  We digitized the velocity curves  of Fig. 8 in \citet{Vasan+2024}                                                                                                                                                                                                                                                                               and obtained the nebular and wind velocity as functions of position along the galactic major axis.The nebular velocity and the wind velocity exhibit each a  large scale shear.  
 We subtracted from each velocity field the corresponding large scale shear, and then removed the remaining mean value. Doing so resulted in two residual velocity fields along the major galactic axis.  We derived the autocorrelation function and the structure function for each. 
   
     Structure functions rather.than   power spectra were employed since the former are more detailed on the smaller and medium spatial scales. They are also  more reliable at treating data at the borders of the data domain \citep{Nestingen-Palm+17}.

 \subsection{The residual nebular velocity along the  galaxy major axis}
  
\begin{figure}[ht!] 
  \centering
   \includegraphics[scale=0.3] {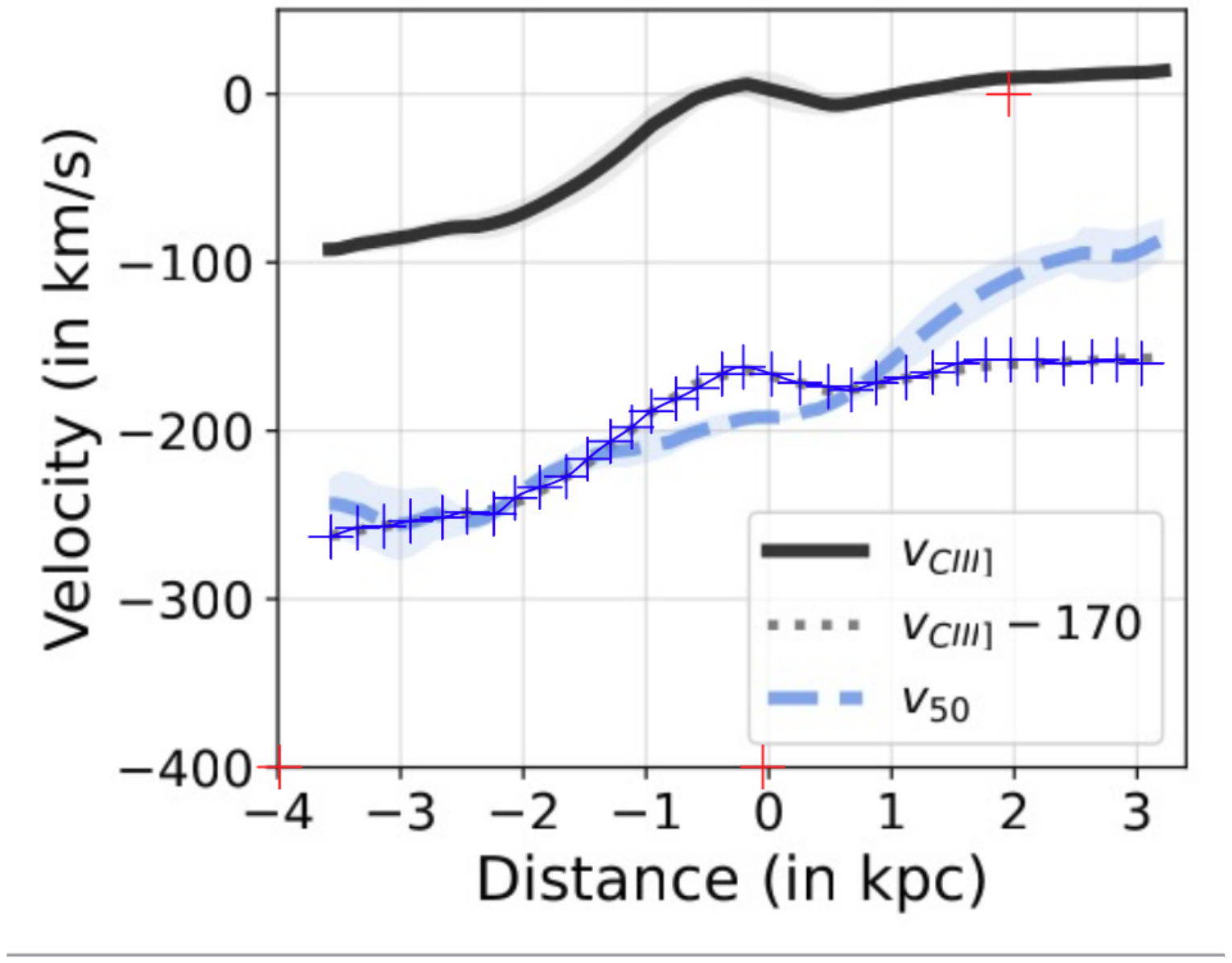} 
  \caption {The marked nebular velocity, offset by 170 km/s, in units of $km/s$ as function of position along the major axis, in units of $kpc$.}
  \label{v_nebdig}
       \end{figure}

We used the Engauge Digitizer Ver.12.1, to mark  the nebular velocity of Fig. 8 in \citet{Vasan+2024}.
The marked observed nebular velocity,   offset by 170 km/s,   is shown in figure \ref{v_nebdig}.

\begin{figure}[ht!] 
  \centering
   \includegraphics[scale=0.4] {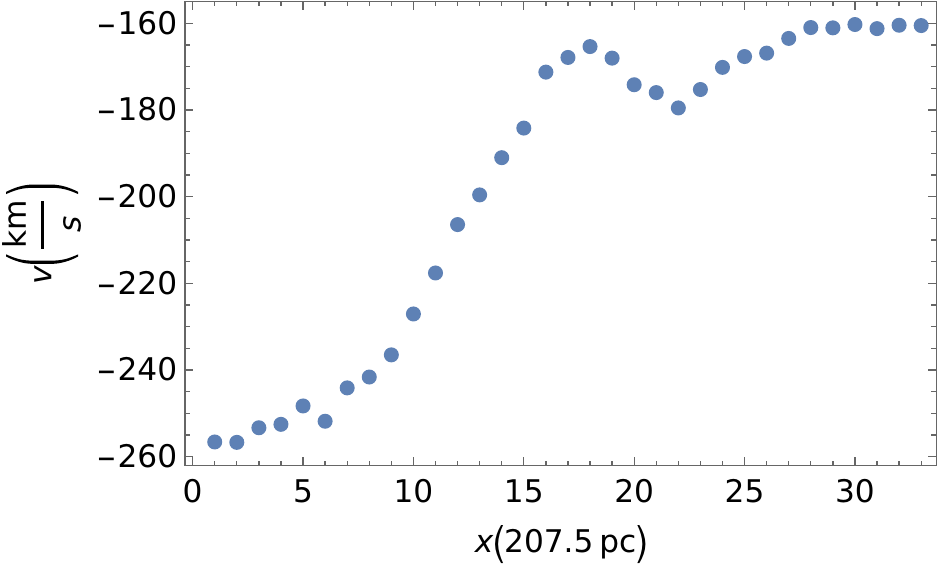 }
  \caption {The  nebular velocity in units of $km/s$ as function of position along the major axis, in units of $207.5pc$.}
  \label{v_neborig}
       \end{figure}

    The Engauge Digitizer yielded the values of the velocity and position at the marked points as seen in figure \ref{v_neborig}.
    
\begin{figure}[ht!] 
  \centering
   \includegraphics[scale=0.4] {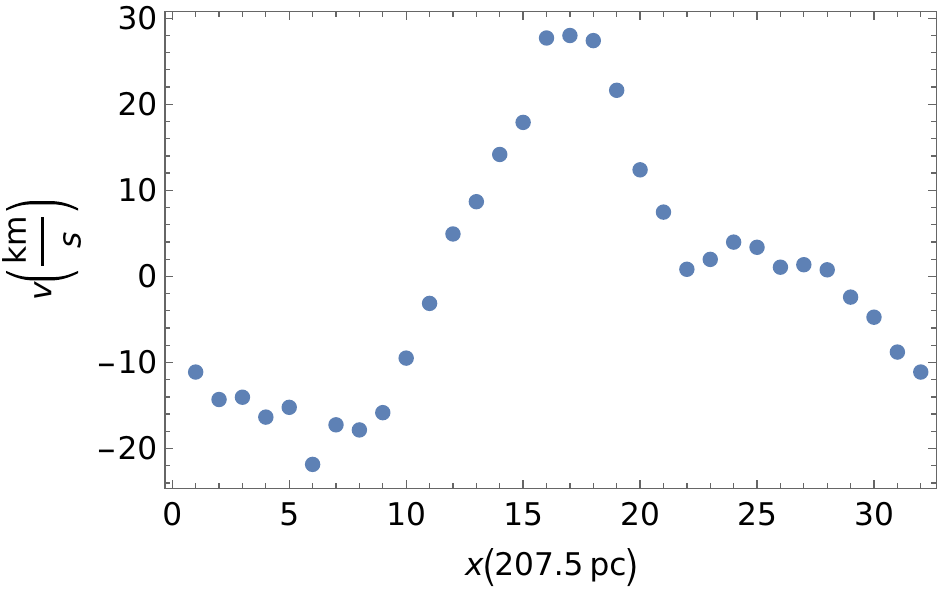 }
  \caption {The residual nebular velocity in units of $km/s$ as function of position along the major axis, in units of $207.5pc$.}
  \label{vneb}
       \end{figure}

 The nebular velocity posses a  large scale shear of $96.2km \ s^{-1}/ ( 6.43 kpc)$.   After subtracting the shear, and the remaining mean value, the residual velocity is obtained and is displayed in Fig.\ref{vneb}.

  \subsection{The residual wind velocity along the  galaxy major axis}
  
\begin{figure}[ht!] 
  \centering
   \includegraphics[scale=0.3] {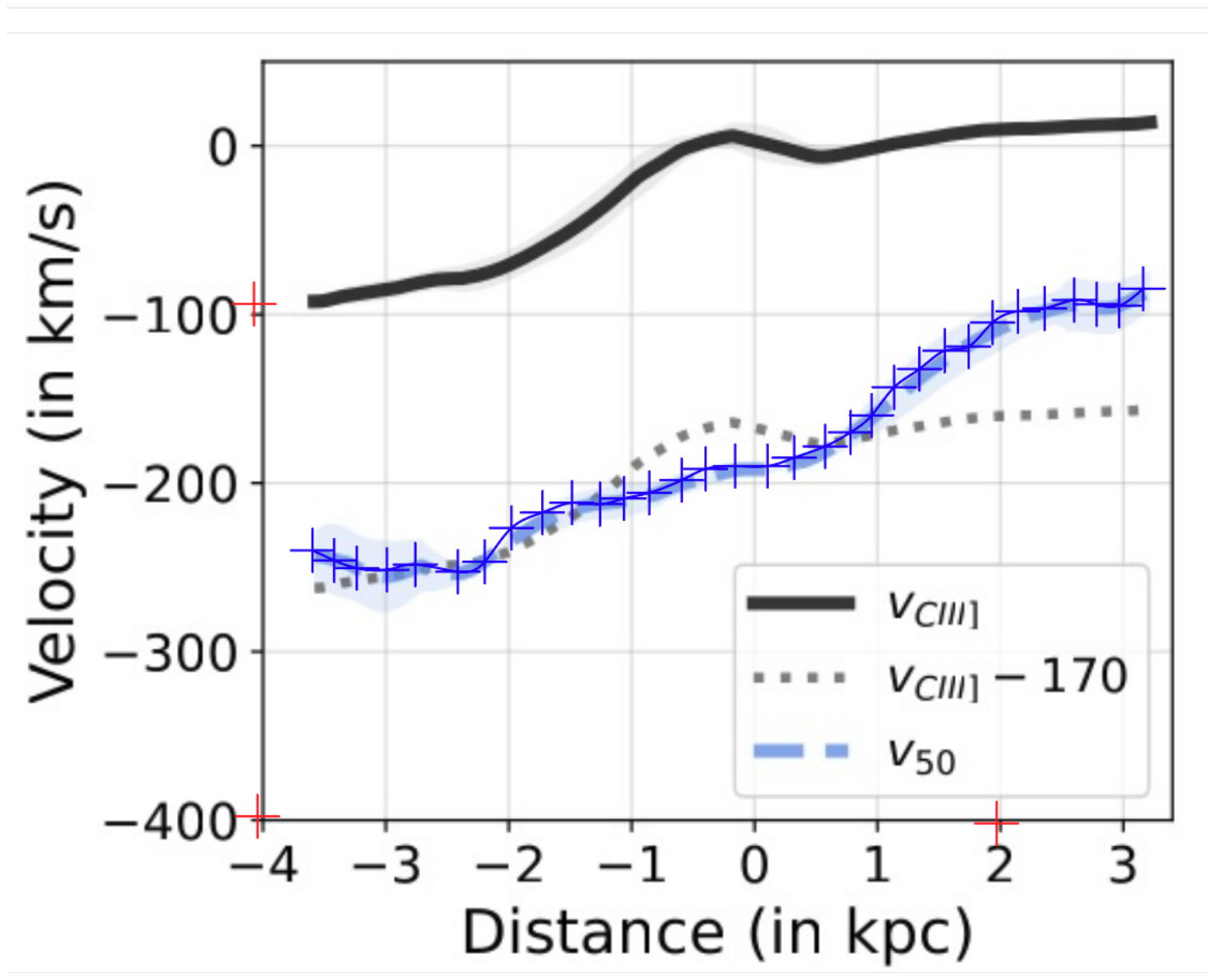} 
  \caption {The marked  wind velocity in units of $km/s$ as function of position along the major axis, in units of $kpc$.}
  \label{v50dig}
       \end{figure}

  We follow similar   steps in treating the wind velocity. The marked digitized wind velocity of Fig. 8 from \citet{Vasan+2024},  is displayed in figure \ref{v50dig}.

\begin{figure}[ht!] 
  \centering
   \includegraphics[scale=0.4] {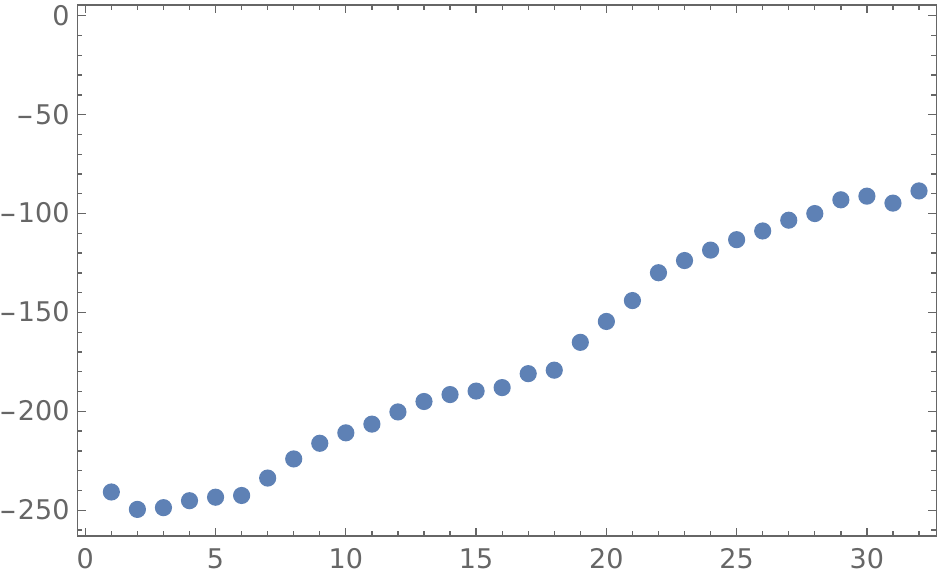 }
  \caption {The   wind velocity in units of $km/s$ as function of position along the major axis, in units of $207.5pc$.}
  \label{v_50orig}
       \end{figure}

 The Engauge Digitizer yielded the values of the velocity  and position of the marked points as seen in  Fig. \ref{v_50orig}.

\begin{figure}[ht!]
  \centering
   \includegraphics[scale=0.4] {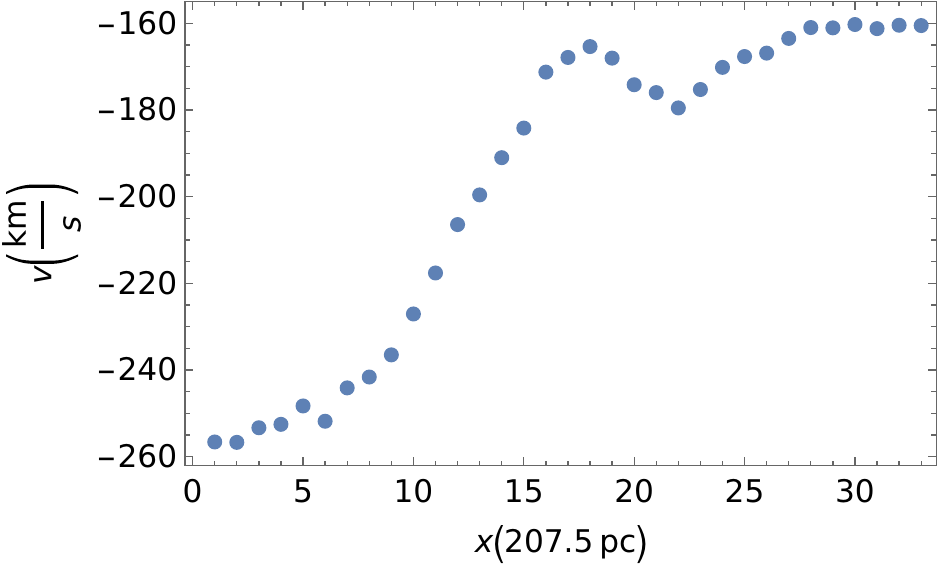}
      \caption{The residual wind velocity in units of $km/s$ as function of position along the major axis, in units of $207.5pc$.} 
             \label{v_50}
                    \end{figure}

    The wind velocity posses a  large scale shear of $152.2 km \ s^{-1}/ ( 6.43 kpc)$.   After subtracting the shear, and the remaining mean value, the residual velocity is obtained and is displayed in Fig.\ref{v_50}.
    
  \section{autocorrelation functions}
  
The one-dimensional autocorrelation   function, $C(x)$, of a residual velocity $v(x)$ (with a zero mean value) is  

\begin{equation}
C(x)= <v(x +x') v(x')>.
\end{equation}
 
The brackets indicate   average over all values of $x'$.  Here $x$ denotes the  spatial lag between   two positions along the  galaxy  major axis. 
                                                                                                                                                                                                                                                                                 
  \subsection{The autocorrelation function of the residual nebular velocity}
\begin{figure}[ht!]
  \centering
 \includegraphics[scale=0.4] {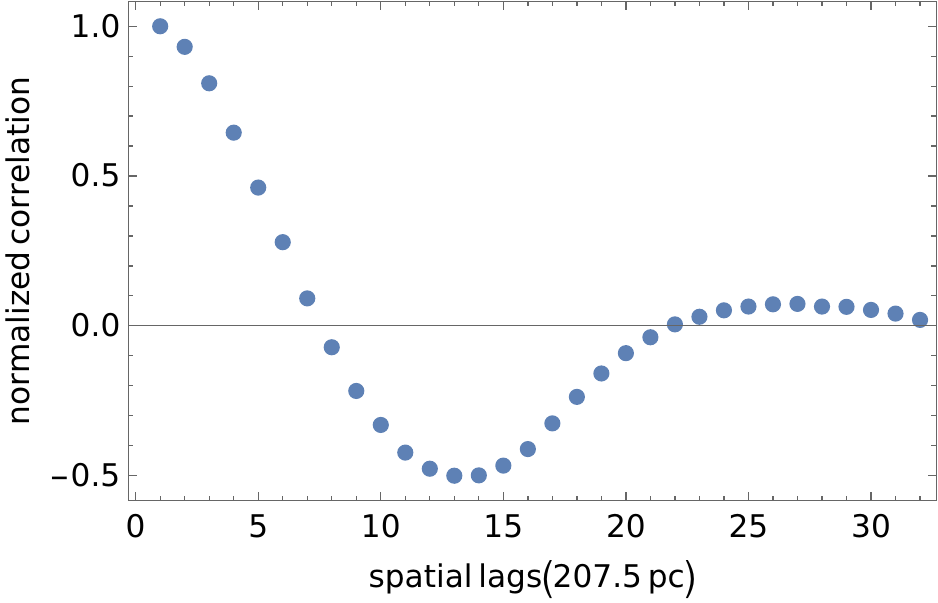 } 
   \caption{The normalized observational
   autocorrelation function of the residual  nebular velocity as function of the spatial lag,  in units $207.5 pc$. }  
     \label{autoc_neb}
     \end{figure}

 The computed observational  normalized autocorrelation function of the  residual nebular velocity  is displayed in Fig, \ref{autoc_neb}. It implies that  the   residual values of the nebular velocity are in fact correlated over large spatial range,  comparable to the size of the major axis.

 \subsection{The autocorrelation function of the residual wind velocity}
 
\begin{figure}[ht!]
 \centering
   \includegraphics[scale=0.4] {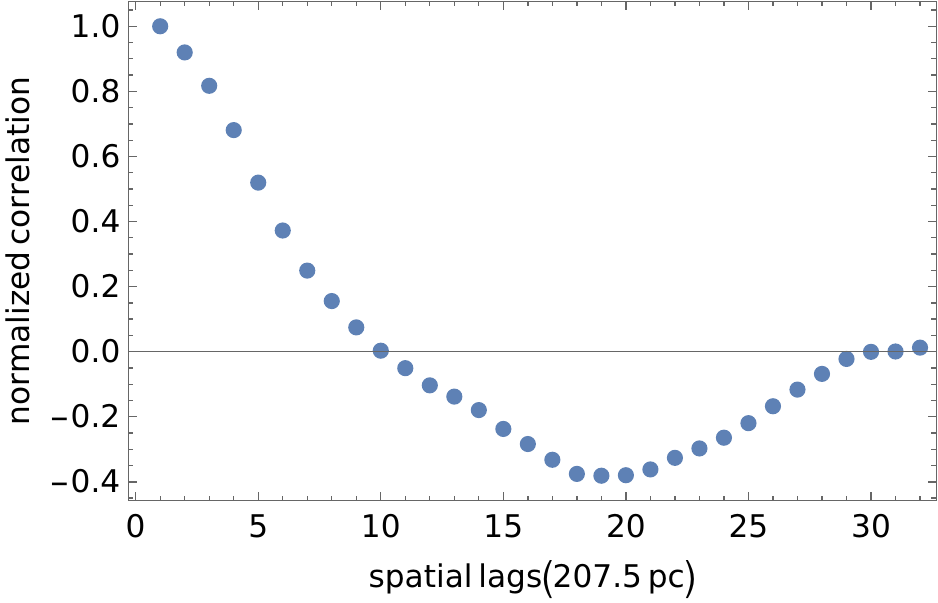 }
   \caption{The normalized observational
   autocorrelation function of the residual  wind velocity as function of the spatial lag,  in units of $207.5 pc$.}  
     \label{autoc_v50}
     \end{figure}

   Fig.\ref {autoc_v50} presents the normalized autocorrelation function of the wind velocity. It exhibits    long range correlation similar to that of the nebular velocity.

 Note that  while the two residual velocity fields displayed in Fig.\ref{vneb} and in Fig.\ref{v_50} appear to be quite different, their autocorrelation are quite similar, implying underlying spatial correlation that are similar.
       Long range  autocorrelation could be  a signature  of turbulence. In  order to test for the existence of turbulence and understand its nature,   we evaluate the structure functions for the two residual velocity fields.  
 
 \vskip 2cm

   \section{structure functions}
   The one-dimensional structure function of a quantity $f(x)$ defined along a  straight line is  
   
   \begin{equation}
   S_f(x)= < \left(f(x' +x) - f(x')\right)^2  >= 2 C_f(0) - 2 C_f(x),
   \end{equation}
  with the lag  $x$ being  the    difference between  two positions. $C_f(x)$ is the auto correlation function of $f(x)$, with zero mean value, defined as
  
  \begin{equation} 
  C_f(x)= <f(x'+x) f(x)>.
  \end{equation} 
  
  In the following, the  computed   structure functions of the observational residual nebular velocity, and of the residual wind velocity are presented,
  
     \subsection{The observational structure  function of the residual nebular velocity}
\begin{figure}[ht!]
    \centering
   \includegraphics[scale=0.4] {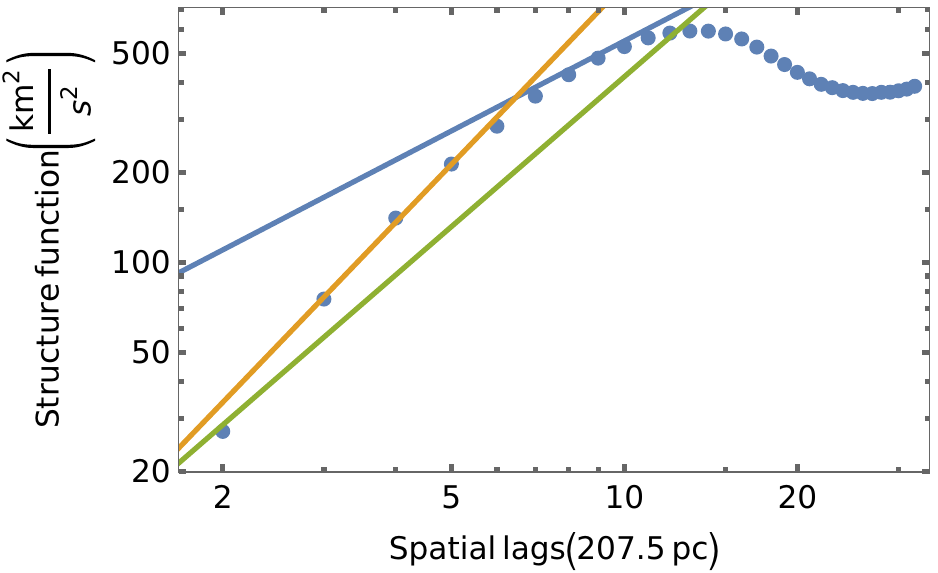 }
    \caption{The observational structure function  of the residual  nebular velocity in units of $(km/s)^2$ as function of the spatial lag, in units of $207.5pc$. The  asymptotes have logarithmic  slopes of 1 and 2. The green line has a logarithmic slope of 5/3.}
     \label{sf_neb}
     \end{figure}

     Fig.\ref{sf_neb} displays the structure function of the observational residual nebular velocity. The  blue   line  has a logarithmic slope of 1. The orange line has a logarithmic slope of 2.  In Appendix A it is shown that the  structure function,  evaluated along a lateral line, of a quantity that is an integral over the line of sight direction, increases the logarithmic slope
by 1 when the lateral lag is  smaller than the effective depth of the turbulent layer. The green line, has a logarithmic slope of 5/3; that of a Kolmogorov turbulence for the lags smaller than the depth.
 
   The structure function at the largest lag 
is  $2 C(0)^2=391 (km/s)^2$. Thus, the one dimensional rms turbulent velocity is $13.9 km/s$. Assuming isotropic turbulence, the three-dimensional turbulent velocity is  $24 km/s$.

     \subsection{The observational structure  functions of the residual wind velocity}
\begin{figure}[ht!]
    \centering
   \includegraphics[scale=0.4] {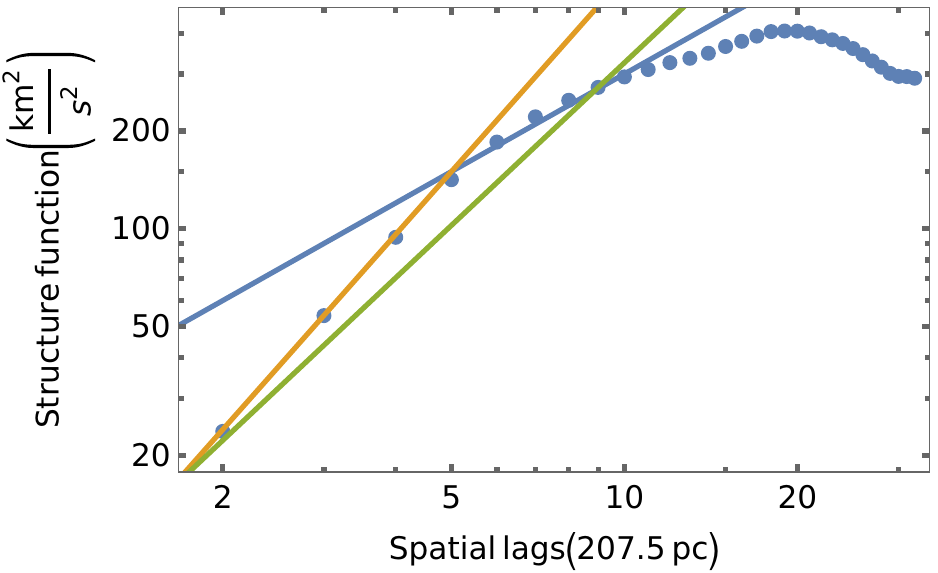 }
   \caption{The observational structure function   
    of the residual wind velocity, in units of $(km/s)^2$ as function of the spatial lag in units of $207.5 pc$. The  asymptotes have logarithmic slopes of 1 and 2.The green line has a logarithmic slope of 5/3.}  
     \label{sf_v50}
     \end{figure}

    Fig.\ref{sf_v50} displays the structure function of the residual wind velocity. The  blue   line  has a logarithmic slope of 1. The orange line has a logarithmic slope of 2. This behavior is similar to that of the structure function of the residual nebular velocity. Here too, the green line has a logarithmic slope of 5/3.
    
      The structure function at the largest lag 
is  $2<C(0)^2=291 (km/s)^2$, implying a one-dimensional   rms turbulent velocity of $12.1 km/s$. Assuming isotropic turbulence, the three-dimensional turbulent velocity is  $21 km/s$.

     \section{discussion}

     \subsection{The nature of the  turbulence}\ 
 The observational structure functions of the two residual velocity fields have, each,  a logarithmic slope equaling 1      on the large spatial scales and  a logarithmic slope of 2 on the small  spatial  scales. This dependence characterizes compressible turbulence with  a one-dimensional power spectrum $\propto k^{-2}$ with $k$ denoting the one -dimensional spatial wave number.  
 
 We obtained in  appendix A that the logarithmic slope of    the structure function equals 1 for spatial lags larger than  $0.547 D$. In   
Fig. \ref{sf_neb} and  Fig. \ref{sf_v50} this logarithmic slope holds for spatial lags $2.822\ kpc\geq  x \geq 1.452\ kpc$. The lower limit is $0.547 D$, where $D$ is the effective depth of the turbulent region.
The upper limit depends on the   value  of the largest spatial lag. In Appendix A the results were obtained for lags that could extend to infinity. Here, when the lag approaches the  largest  value, the asymptotic logarithmic slope of 1 is no longer valid.

     The compressible power spectrum is  steeper than the Kolmogorov power spectrum, which corresponds to subsonic incompressible turbulence
with a one-dimensional  power law exponent of  $-5/3$ and structure function with logarithmic slope of 2/3 and 5/3 for the large and small scales, respectively.
In Fig.  \ref{sf_neb}   and in Fig.  \ref{sf_v50}, we plotted also the a line with a logarithmic slope of 5/3 corresponding to Kolmogorov turbulence for the small spatial lags. 
It is seen that the logarithmic slope of compressible turbulence is favored. 

A compressible power spectrum was derived by \cite{Burgers48} describing a hierarchy of shocks in compressible gas. Compressible turbulence power spectra were observed in HI  intensity maps in the Milky Way (MW) galaxy \citep{Green93}and in the SMC \citep{Stanimirovic+99}. This power spectrum has been observed also  in molecular clouds \citep{Larson81,Leung+82,Dame+86}; in the   HII region Sharpless~142  \citep{Roy+Joncas85}. It has been found in a shocked cloud near the Milky Way galaxy center \citep{Contini+Goldman2011}, and recently in the Gamma ray emission from the large Magellanic Cloud \citep{Besserglik+Goldman2021}. It has been   obtained also in numerical simulations e.g.  \citep{Passot+88,Vazquez-Semadeni+97, Kritsuk+2007,Federrath+2021}.  
 
           The steeper slope signals that (unlike in the Kolmogorov spectrum)
the rate of energy transfer in the turbulence cascade is not constant but  decreases with increasing wavenumber. This is   expected in a compressible turbulence since part of the energy at a given wavenumber in the cascade, is diverted to compression of the gas. Indeed, a theoretical derivation of the compressible turbulence power spectrum based on this argument has been obtained  \citep{Goldman2021a}. 

The three-dimensional rms turbulent velocities estimated in section 4.are supersonic, in line with the turbulence being a compressible one.

   The turbulence timescale of  the largest scale eddies  is  $\sim L/v_0  \sim 200$ Myr where $L=6.44 \ kpc$ is the largest spatial scale and $v_0$ is the turbulent velocity on this scale. This timescale represents the eddy correlation time on the largest    spatial scale and therefore a lower bound on the time span over which the turbulence was  created. This time span is an order of magnitude larger than the age of the young   stars and  the outflowing wind.  Thus, the turbulence is older than the young stars and the wind that was created by the latter.  The timescales of the large scale shears are about 20 Myr.  The emerging picture is that 
the young stars as well as the wind and the shears were formed on the background of the turbulent interstellar gas.

The autocorrelation functions reveal that the two velocity fields are correlated over a scale of $\sim 6.44\ kpc$.The generating source  of this large scale turbulence, must itself be correlated over the largest scale of the turbulence, rather than be a collection of smaller scale  sources. The probable source is therefore a merger   or a close tidal interaction with a smaller galaxy. If the resulting flows are supersonic, the turbulence generated by instabilities in these flows will be supersonic compressible turbulence e.g. \citep{Sparre+2022}, as seems to be in the present case.

  \subsection{The effective depth of the turbulent region}
  
    The  emitted photons that  determine the velocities originate from different depths along the line of sight.  
  The issue of power spectra and structure functions  of quantities which  are the result of integration along the line-of-sight has been addressed by e.g.  \citep{Stutzki+98, Goldman2000, Goldman2021b, Lazarian+Pogosyan2000,   Miville+2003a}. These authors concluded that when the lateral spatial scale is smaller than the depth of the turbulent layer, the logarithmic slope of the power spectrum 	 steepens   by $-1$ compared to its value when the lateral scale is large compared to the depth. The logarithmic slope of the structure function increases by 1. This behavior was indeed  revealed  in  observational power spectra  of Galactic and extra Galactic turbulence ( e.g. 
\citet{elmegreen+2001},\citet{ Block+2010}, \citet {Miville+2003b} ) and in solar photospheric turbulence \citep{Abramenko+Yurchyshyn2020}).

   In Appendix A,  the theoretical  structure function of a quantity that is the result of integration along the line of sight, is obtained. For the specific case of compressible turbulence we found that   $D=1.83  x_{tr}$, where $D$ is the effective depth of the turbulent layer and $x_{tr}$ is the observational lag  marking the slopes transition in the structure function. 
The effective depth is the depth of a layer with depth independent of the lateral coordinate, that would yield the observational structure function.  

   From Fig.\ref{sf_neb} the  observational transition lag for the nebular  velocity  
       $(1.31\pm 0.04) kpc$   yielding an effective depth of $(2.4\pm 0.07) kpc$. The observational transition  lag of the wind   velocity structure function,  from Fig.\ref{sf_v50} is $(1.06\pm 0.04) kpc$ implying an effective depth of $(1.9. \pm 0.07) kpc$. 
       
 The two estimates are close but different. This may occur if the emission lines  that determine the velocities emanate from different depths.  The values of the effective depth are consistent with the conclusion of    \citet{Vasan+2024} that the wind that, in a spherical shell model,  is enclosed in a shell of width $\sim 2\ kpc$. Our results indicate that the nebular gas disk is quite a thick one, this is in line with observations and simulations of high z galaxies, see e.g. \citep{Genzel+2006, Genzel+2008}.

\subsection{Power spectra of the residual velocity fields } 

\begin{figure}[ht!]
    \centering
   \includegraphics[scale=0.4] {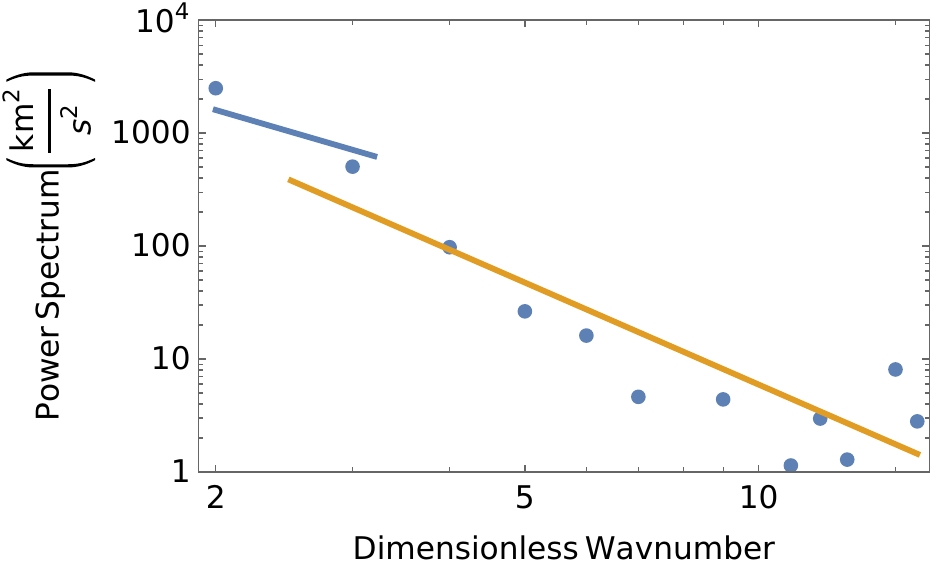}
   \caption{The power spectrum    
    of the residual nebulae velocity, in units of $(km/s)^2$ as function of the dimensionless wavenumber. The blue line has  has a logarithmic slope of  -2 and the orange line has a logarithmic slope of -3.}  
     \label{nebps}
     \end{figure}

\begin{figure}[ht!]
    \centering
   \includegraphics[scale=0.4] {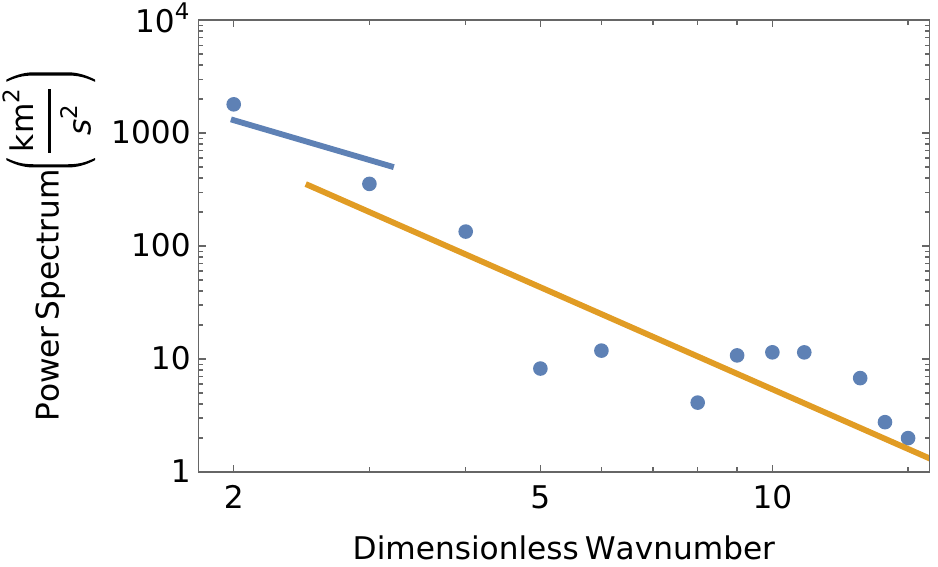 }
   \caption{The power spectrum    
    of the residual wind velocity, in units of $(km/s)^2$ as function of the dimensionless wavenumber. The blue line has  has a logarithmic slope of  -2 and the orange line has a logarithmic slope of -3.}  
     \label{v50ps} 
     \end{figure}

To complement the present study we obtain the observational power spectra of the residual velocity fields.
The power spectra are the squared absolute value of the discrete Fourier transform of the residual velocity fields. The power spectra of the nebular velocity and the wind velocity are presented in Fig.\ref{nebps}and Fig. \ref{v50ps}, respectively.  The plots also show straight lines with logarithmic slopes of -2 and -3. 
For both velocity fields the small wavenumber (large spatial scale) region is not evident. 

The power spectra in the figures  are functions of the dimensionless wavenumber $q= k_x/k_0$ where $k_0= \frac{2\pi}{L}$ is the smallest wavenumber corresponding to the largest spatial scale $L$.
In Appendix B, it was obtained that   the power spectra are functions of  $\mu= \frac{k_x D}{2}$ and that  the transition value between the two regimes is at $\mu_{tr}=1.54$.Thus, the transition wave number is $k_{x, tr}= \mu_{tr}\frac{2}{D}$ implying that the transition dimensionless  wavemeter is
  
\begin{equation}
q_{tr}= \mu_{tr} \frac{2}{D}\frac{L}{2\pi}=1.54 \frac{L}{\pi D}
\end{equation}
 
 Employing $L=6.44 kpc$ and the depth values $D= 2.4,\ 1.9\ kpc$ for the nebular and wind velocity fields respectively, results in $q_{tr}=1.31,\ 1.66$, respectively. This explains why    Fig.\ref{nebps} and Fig.\ref{v50ps} do not show the small wavenumber region. Thus,  the advantage of using structure functions rather than power spectra, in the present case, is clear. If the   depth was not that large, then power spectra would be a useful tool.

     \newpage
     \section*{Acknowledgment}
I thank  the referee for helpful comments and the Research Authority of Afeka College for support.

   \appendix

  \section{A. The theoretical structure function of a quantity that is the result of integration along the line of sight}

       Consider a function $f(x)$ where $x$ is a straight line in a lateral direction.  and is  an integral  along the line of sight:  
      
      \begin{equation}
      f(x) = \int_0^D g(x, z) dz.
      \end{equation}
      Here, $z$ is the line of sight coordinate and  $D$ the depth. A plane parallel geometry is assumed for simplicity.

  The autocorrelation function of $f(x)$ is: 
  
  \begin{eqnarray}  
  C_f (x)= <f(x+x') f(x)> =\\
   \int_0^D\int_0^D<g(x', z) g(x+x', z') > dz dz'=\nonumber\\
   \int_0^D\int_0^D C_g(x, z-z')dz dz'.\nonumber
    \end{eqnarray} ,
  
 The autocorrelation function  $C_g(x, z-z')$ can be expressed by the two-dimensional power spectrum, $P_2(k_x, k_z)$,
 \begin{eqnarray}
  \hskip -1.5cm C_g(x, z-z')= \int_{-\infty}^{\infty} \int _{-\infty}^{\infty}e^{i( k_x x +k_z ( z-z'))   }\\
  \nonumber
   P_2(k_x, k_z) dk_x dk_z.  
    \end{eqnarray}
  leading to
   \begin{equation}
    C_f (x)=\int_{-\infty}^{\infty} \int _{-\infty}^{\infty}e^{i k_x x}    P_2(k_x, k_z) \frac{  \sin^2  \left( k_z D/2  \right)}       {\left(   k_z D/2  \right)^2 } dk_x dk_z.
    \end{equation}
    
 from equations (2) and (A4) follows the expression for the   structure function
 
\begin{eqnarray}  
 S_f(x, D) 
 \propto  \int_0^{\infty} \int_0^{\infty} \sin^2(k_x x/2)  \frac{\sin^2  \left( k_z D/2  \right)}{  \left(   k_z D/2  \right)^2}\\
 \nonumber 
  P_2(k_x, k_z) dk_x dk_z. 
\end{eqnarray} 

 In the case of a turbulence with a one-dimensional power spectrum  which is a power law with index $-m$, the two dimensional power spectrum is
 \begin{equation}
 P_2(k_x, k_z) \propto   \left(k_x^2 +   k_z^2\right)^{-(m+1)/2 .}\ \ \  \ \ \  
 \end{equation}
 resulting in
 
   \begin{eqnarray}  
 S_f(x, D) \propto  \int_0^{\infty} \int_0^{\infty} \sin^2(k_x x/2)  \frac{\sin^2  \left( k_z D/2  \right)}{  \left(   k_z D/2  \right)^2}\\
 \nonumber
  \left(k_x^2 +   k_z^2\right)^{-(m+1)/2 } dk_x dk_z. 
\end{eqnarray} 

It is convenient  to define the dimensionless variables
\begin{equation}
 \eta= k_z D/2. \ \ \ \ \  \ \ ; \ \ \ \mu= k_x D/2.  \nonumber
\end{equation}

     \begin{figure}[ht!]
    \centering
   \includegraphics[scale=0.4] {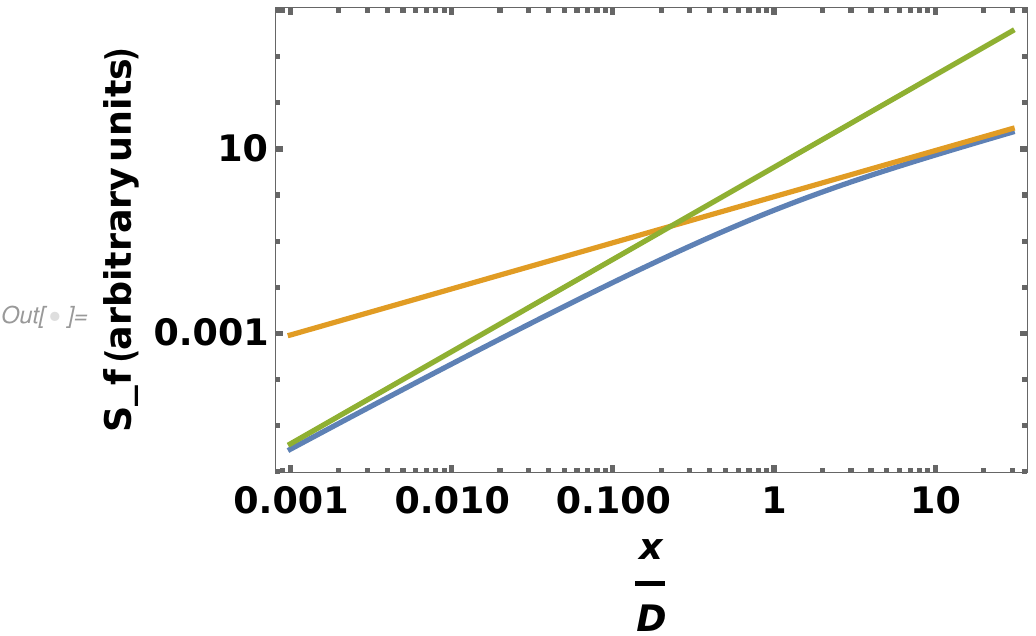 }
   \caption{Theoretical 
    structure function  of  a quantity, which is an integral  
      along the line of sight, as function of$x/D$. The   orange line has logarithmic slopes equaling 1 and the green line has a logarithmic slope of 2.}  
     \label{sf_th1}
     \end{figure} 
     
      \begin{figure}[ht!]
    \centering
   \includegraphics[scale=0.4] {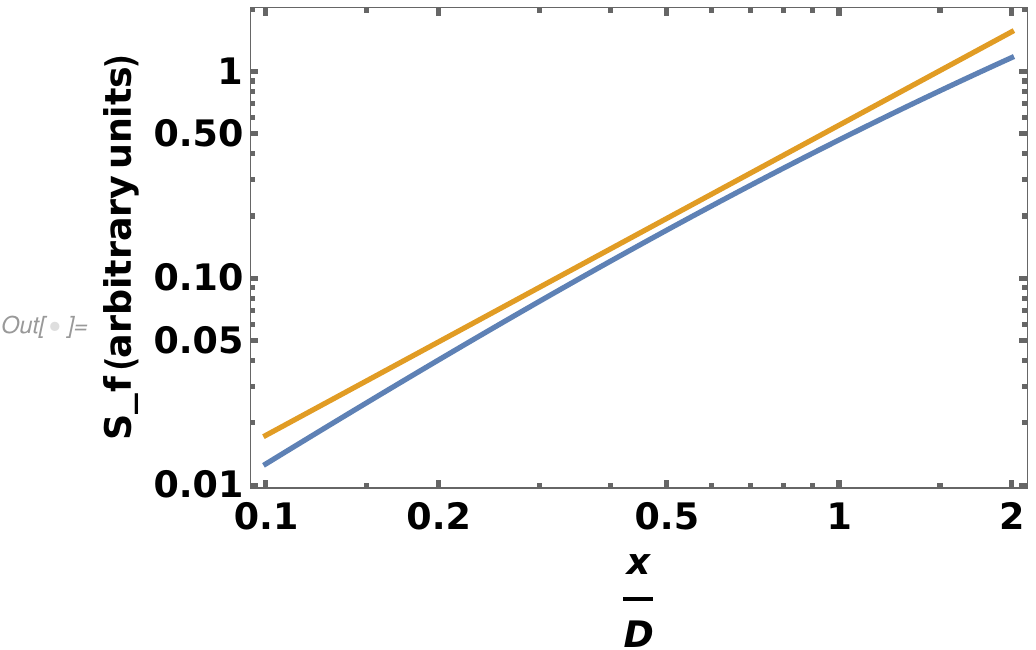 }
   \caption{Theoretical orange
    structure function   
    of  a quantity, which is an integral along the line of sight, as function of$x/D$. The straight line has a logarithmic slope of 1.5.}  
     \label{sf_th2}
     \end{figure}   
     
     The structure function of equation (A8) can be expressed as       
         \begin{equation}                        
  S_f(x ,D)
   \propto \int_0^{\infty}I(\mu)\sin^2\left( \mu x/D  \right)    d\mu.
      \end{equation}
  
 where  $I (\mu)$ is
 \begin{equation}
        I(\mu )=   \int_0^{\infty}  \left(\mu^2  +  \eta^2\right)^{-(m+1)/2} \frac{ \sin^2 \eta}{     \eta^2} d\eta.
   \end{equation}

Equation (A10)    implies  that the structure function  argument is $x/D$. 
Also, inspection of   equations (A10) and (A11) reveals that for $x<<D$ the structure function is proportional to $x^m$ while for $x>>D$
it is proportional to  $x^{m -1}$ .
 
 A numerical solution for the case of $m=2$ is presented in Fig.\ref{sf_th1} together with  power laws with exponents 1 and 2. 
  In order to find the value of $x_{tr}/D$, where  $x_{tr}$ denotes the transition lag between the two slopes,   a power law with exponent of 1.5 is plotted in  Fig.\ref{sf_th2} together with the structure function. The value of $x_{tr}/D$ is taken to be   the value for which the logarithmic
 slope of the structure function equals 1.5, i.e. where the straight line  is tangent to the structure function.
  The result is $x_{tr}/D \sim 0.547$; thus   $D \sim 1.83 x_{tr}$.

  \section{B. The theoretical power spectrum of a quantity that is the result of integration along the line of sight} 

The 1D power spectrum $P_f(k_x)$ is by definition

\begin{equation}
P_f(k_x)= \int_{-\infty}^{\infty} e^{-ik_x} C_f(x) dx
\end{equation}
\\
From equations (A4) and (A6) follows

\begin{equation}
P_f(k_x) \propto \int_{-\infty}^{\infty}\left(k_x^2 +   k_z^2\right)^{-(m+1)/2 }  \frac{\sin^2  \left( k_z D/2  \right)}{  \left(   k_z D/2  \right)^2} dk_z
\end{equation} 

Using Equations (A9) and A(11) we obtain
\begin{equation}
P_f(k_x) = P_f(\mu)= I(\mu)
\end{equation}
 
\begin{figure}[ht!]
 \centering
   \includegraphics[scale=0.4] {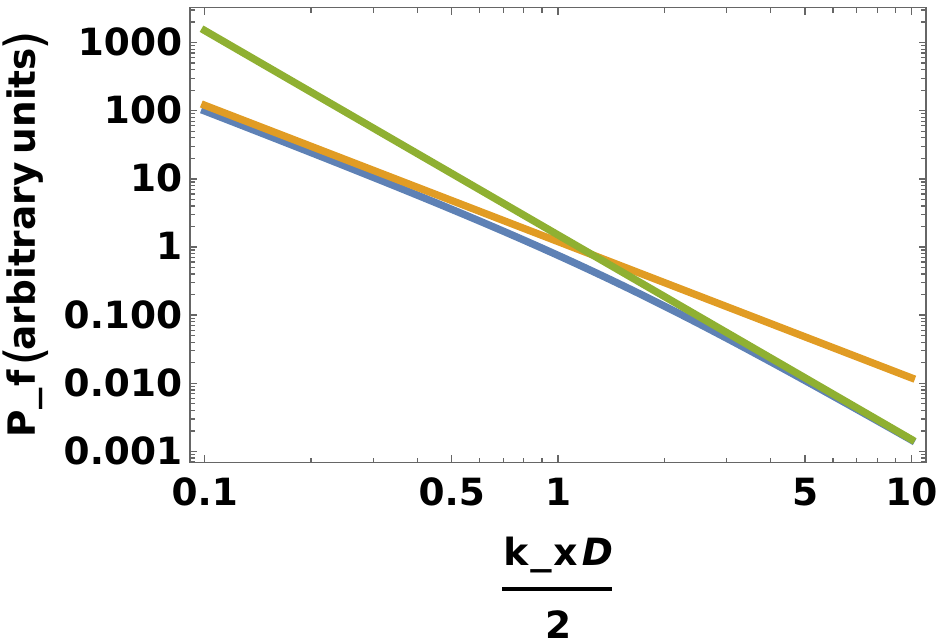 }
   \caption{Theoretical 
power spectrum, of a quantity which is an integral  
      along the line of sight, as function of $k_x D/2$. The orange line has a logarithmic slopes equaling -2 and the green line has logarithmic slope of -3.}  
 \label{ps_th1}
     \end{figure}

Thus, the power spectrum is a function of $\mu$. Inspection of equations (A11) and  (B3) reveals that for 
$\mu<<1$, $P_f( k_x)\propto k_x^{-m}$, while for $ \mu>>1$,  $ P_f( k_x)\propto k_x^{-m-1} $. For the present case of $m=2$ the power spectrum is plotted in Fig.\ref{ps_th1}. Plotted also are the asymptotes 
with logarithmic slopes of -2 and -3.

\begin{figure}[ht!]
 \centering
   \includegraphics[scale=0.4] {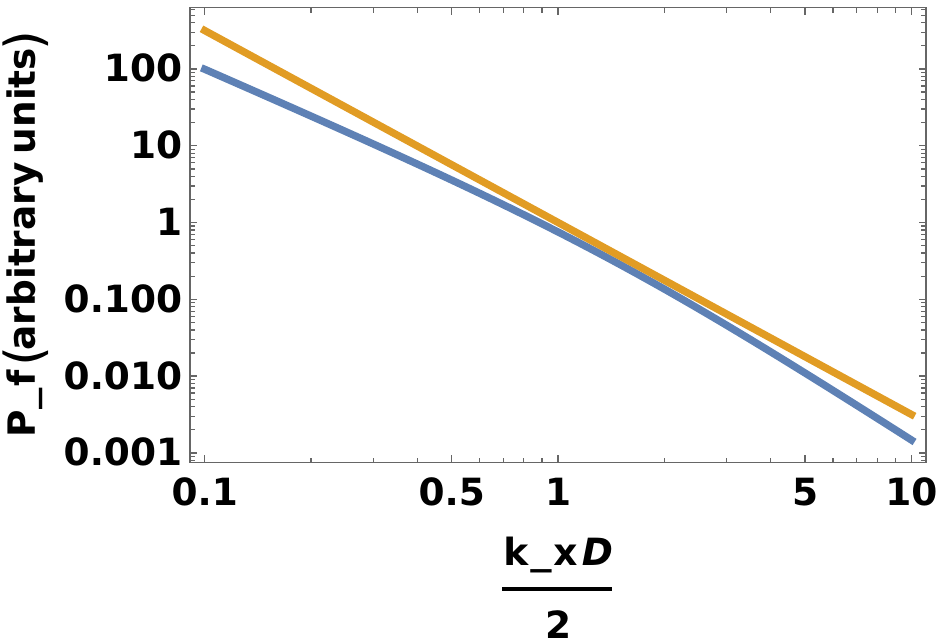 }
   \caption{Theoretical 
power spectrum, of a quantity which is an integral  
      along the line of sight, as function of $k_x D/2$. The straight line has a logarithmic slope equaling -2.5.}  
 \label{ps_th2}
     \end{figure}

In order to find the transition value $\mu_{tr}$ a line with a logarithmic slope of -2.5 is plotted and the value of $\mu_{tr}$ is taken to be the tangent point to the power spectrum. From Fig.\ref{ps_th2}
we find that $\mu_{tr}=1.54$.

          \end{document}